\documentclass[prl,twocolumn]{revtex4}
\usepackage{graphicx}

\usepackage{epstopdf} \DeclareGraphicsExtensions{.eps, .pdf}

\begin{document}

\def\K{{\bf{K}}}
\def\Q{{\bf{Q}}}
\def\X{{\bf{X}}}
\def\Gbar{\bar{G}}
\def\tk{\tilde{\bf{k}}}
\def\k{{\bf{k}}}
\def\q{{\bf{q}}}
\def\x{{\bf{x}}}
\def\y{{\bf{y}}}

\title{The isotope effect in the Hubbard  model with local phonons}
\author{Alexandru Macridin and M.\ Jarrell}
\address{
University of Cincinnati, Cincinnati, Ohio, 45221, USA}

\date{\today}

\begin{abstract}
The isotope effect (IE) in the two-dimensional Hubbard model with Holstein phonons is studied
using the dynamical cluster approximation with quantum Monte Carlo. 
At small  electron-phonon (EP) coupling the IE is negligible. For larger  EP coupling
there is a large and positive IE on the superconducting temperature that decreases with increasing doping.
A significant IE also appears in the low-energy density of states, kinetic energy and charge excitation spectrum.   A negligible IE is found in the pseudogap and antiferromagnetic (AF) properties at small doping whereas the AF susceptibility at intermediate doping increases with decreasing phonon frequency $\omega_0$.  This IE stems from increased polaronic effects with decreasing  $\omega_0$.  A larger IE at smaller doping  occurs due to stronger polaronic effects
determined by the interplay of the EP interaction with  stronger AF correlations. The IE of the 
Hubbard-Holstein model exhibits many similarities with the IE measured in cuprate superconductors.

\end{abstract}

\maketitle


{\em{Introduction.}} Isotope effect (IE) measurements \cite{IE} played 
a significant role in establishing a phonon-mediated mechanism in conventional 
superconductors.  One of the greatest successes of 
the Bardeen-Cooper-Schrieffer (BCS) theory of superconductivity \cite{BCS} was  
the ability to explain the IE experimental data.  However,  the IE in the
high T$_c$ materials \cite{IEcuprates,IEzhao,Khasanov} is unusual.  
At optimal doping the IE is small, leading many researchers to argue that 
phonons do not play an important role in the mechanism of high T$_c$ superconductivity and
consequently that its origin must be purely electronic. Nevertheless, the very large 
IE at small doping clearly shows that phonons play an important role in the physics of 
the cuprates. 

Some investigators have proposed the unusually large IE at small doping may be caused by the vicinity
to a quantum critical point, defining a crossover from three-dimensional to 
two-dimensional (2D) physics as the doping decreases \cite{ie_qcp}.  Other 
studies claim that the IE in cuprates can be attributed to polaron formation \cite{keller_polaron}. 
However, most theoretical investigations of the IE in models for high T$_c$ 
superconductors \cite{keller_polaron, ie_dwave,zhao2} start with approximations 
that assume a BCS-type mechanism for $d$-wave pairing and then study the effect of 
phonon characteristic energy on the superconducting properties.  In contrast, the present
study employs the dynamical cluster approximation (DCA) \cite{dca,maier:rev}
and treats, on equal footing, both the EP and electron-electron interactions in 
a 2D system in the form of the Hubbard-Holstein (HH) model. 
 In previous work employing the DCA \cite{jarrell:sc}, the 
single-band Hubbard model, a generally accepted paradigm for the low energy physics of cuprates
 \cite{anderson,zhang_rice},
shows $d$-wave superconductivity driven by strong 
electronic correlations. Furthermore, when combined with EP coupling, the superconducting temperature
is significantly influenced by polaronic effects which  are enhanced in the 
presence of antiferromagnetic (AF) correlations \cite{macridin_hh,macridin_phonon} inherent in the underlying
Hubbard model. Here, by polaronic effects we understand  renormalization of the single-particle propagator,
i.e.  reduction of quasiparticle (QP) weight and charge carrier mobility  due to EP coupling, and not 
necessarily the formation of small polarons or a narrow polaronic band.
Thus far no investigation of the IE in systems with  both strong electronic 
correlations and EP interaction has been conducted with cluster mean-field methods.


For conventional superconductors the BCS theory predicts that the  isotope coefficient 
$\alpha= -\frac{d\ln T_c}{2d\ln \omega_0}=0.5$, where T$_c$ is the superconducting 
temperature and $\omega_0$ the phonon frequency. The HH model with low frequency 
phonons, displays a positive and unusual IE.  For small EP coupling the IE 
is small ($\alpha \ll 0.5$)  while for values of EP  coupling large enough 
to drive the system in to a regime with strong polaronic effects $\alpha$ can 
be very large ($\alpha \gg 0.5$) and  $\alpha$  decreases with increasing  
doping. The IE  in the HH model is directly related with the enhancement 
of polaronic effects with decreasing  $\omega_0$, 
as seen in the dependence of the single-particle quantities and charge susceptibility
on changes in $\omega_0$.  A larger 
IE at smaller doping is due to increased polaronic 
effects in the presence of stronger AF correlations as described in a 
previous investigation \cite{macridin_hh}.  
The pseudogap and AF properties display a negligible IE at 
small doping, whereas the AF susceptibility at intermediate doping increases with decreasing  
$\omega_0$. 
The size, sign and doping dependence of the IE in the HH model
is  similar to  experimental data in the 
cuprates \cite{IEcuprates, ie_pg, ie_af}.

{\em{Formalism.}} 
The HH Hamiltonian reads
\begin{eqnarray}
H&=&-t \sum_{\langle ij\rangle\sigma}\ \left(c^\dagger_{i\sigma} c_{j\sigma} +
c^\dagger_{j\sigma} c_{i\sigma}\right) + U \sum_i n_{i\uparrow} n_{i\downarrow}
\label{Eq:Ham} \nonumber 
\\
&+& \sum_i \frac{p_i^2}{2M} + \frac{1}{2} M \omega_0^2x_i^2 + g  n_i x_i.
\end{eqnarray}
\noindent Here $t$ is the nearest-neighbor hopping, $U$ is the on-site Coulomb repulsion between electrons, 
$\omega_0$ is the energy of the dispersionless optical phonons and $g$ is the on-site EP coupling.  The 
dimensionless EP coupling is defined as $\lambda=2 g^2/(2M\omega_0^2 W)$ and represents the ratio of the 
single-electron lattice deformation energy $E_p=g^2/(2M\omega_0^2)$ to half of the electronic bandwidth  $W/2=4t$.  The 
IE results from the isotopic change of the ions atomic mass $M$ and, since $\omega_0 \propto M^{-\frac{1}{2}}$,
$\lambda$ is kept fixed and $\omega_0$ varies.  Results are presented for small phonon frequency $\omega_0 \le 0.3t$ 
and for $U=8t$, generally accepted or canonical parameter regimes for 
studies of this type and in a range that may be
appropriate for the cuprates \cite{w_cuprates}.

The DCA \cite{dca,maier:rev}, a cluster mean-field theory that maps the original lattice model onto a
periodic cluster of size $N_c=L_c^2$ embedded in a self-consistent
host, is used to study the HH Hamiltonian of (\ref{Eq:Ham}). 
Correlations up to a range $L_c$ are treated explicitly, while
those at longer length scales are described at the mean-field level.
With increasing cluster size, the DCA systematically interpolates
between the single-site mean-field result and the exact result, while
remaining in the thermodynamic limit.  Cluster mean-field calculations on 
the simple Hubbard model~\cite{maier:rev} or Hubbard model with local 
phonons \cite{macridin_hh,macridin_phonon} have been successful at
capturing many key features associated with cuprate materials 
including a Mott gap and strong AF correlations, 
$d$-wave superconductivity and the pseudogap phenomenon.

A quantum Monte Carlo (QMC)
algorithm~\cite{jarrell:dca} modified to perform the sum over both the
discrete field used to decouple the Hubbard repulsion and the
phonon field $x$ serves as the cluster solver in the DCA.  
The present calculations are restricted to clusters of size
$N_c=4$. Low temperature results for larger clusters are inaccessible due to
the sign problem in the QMC calculation.  The Maximum Entropy method~\cite{jarrell:mem} 
is employed to calculate the real frequency spectra.

The superconducting properties are analyzed using the pairing 
matrix $M=\Gamma \chi_0$.  According to the Bethe-Salpeter equation 
\begin{eqnarray}
\label{eq:bs}
\chi=\chi_0+\chi_0 \Gamma \chi=\chi_0(1-M)^{-1},
\end{eqnarray} 
the instability in the two-particle pairing propagator, 
$\chi$, indicating a possible transition to a superconducting state, occurs 
when the leading eigenvalue of $M$ reaches one. The corresponding 
eigenvector $\Phi$ determines the pairing symmetry.  $M$  is the product of  
the pairing interaction vertex $\Gamma$ and the bare bubble $\chi_0=G G$.  
$\Gamma$ can be seen as an effective paring interaction; $\chi_0$ is 
a convolution of the dressed single-particle propagators.  The effect of  EP coupling and  $\omega_0$ on 
these two quantities can be studied by looking at 
$V_d=\sum_{K,i\omega;K',i\omega'} \Phi_d(K,i\omega) \Gamma(K,i\omega;K',i\omega')\Phi_d(K',i\omega') $ 
and $P_{d}=\sum_{K,i\omega} \Phi^2_d(K,i\omega)\chi_0(K,i\omega)$. They 
are the respective projections of the interaction vertex $\Gamma$ and $\chi_0$
on the subspace spanned by the $d$-wave eigenvector $\Phi$, as discussed in 
Ref \cite{maier:pairprb}.

\begin{figure}[t]
\begin{center}
\includegraphics*[width=3.3in]{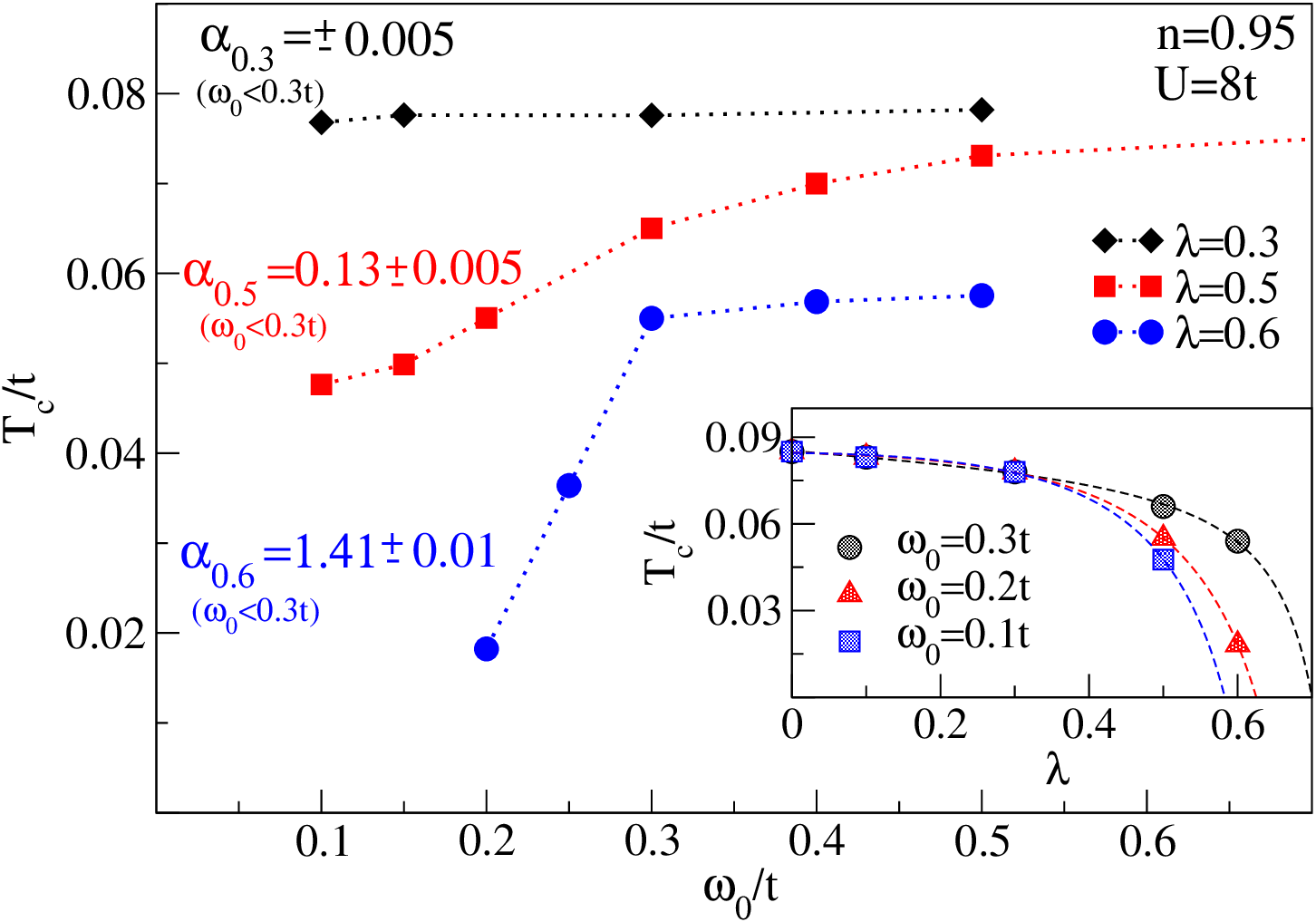}
\caption{(color online) $T_c$ versus  $\omega_0$ at $5\%$ doping for
 $\lambda=0.3, 0.5, 0.6$. Inset:  $T_c$ 
versus $\lambda$ for  $\omega_0=0.3t,0.2t,0.1t$.  The transition to the polaronic regime
 ($0.5 \lesssim  \lambda \lesssim 0.7$) is 
accompanied by an abrupt increase of the IE coefficient $\alpha$. }
\label{fig:ISO}
\end{center}
\end{figure}

{\em{Results.}}  
Previous DCA calculations \cite{macridin_hh} show 
that at small doping  large changes occur in the system properties once 
$\lambda > \lambda_c \approx 0.5$.  The crossover (CRO) region  defining the transition from
large to small polaron regime, $0.5 \lesssim \lambda \lesssim 0.7$ \cite{lambdac}
is characterized by strong polaronic  effects, a suppressed but finite $T_c$ and a large IE effect.  
Outside the CRO region, for small EP coupling
a negligible IE effect is found, whereas for $\lambda>0.7$,  $T_c$ vanishes 
due to formation of small and heavy polarons.
The IE for different values of $\lambda$ is shown in Fig.~\ref{fig:ISO}.
When $\lambda=0.3$,  $T_c$ is nearly 
independent of  $\omega_0$, and just slightly smaller than  
$T_c$ of the Hubbard model without phonons (see the inset).   At 
$\lambda=0.5$  we find a significant reduction of $T_c$ with 
decreasing phonon frequency $\omega_0$ although the IE coefficient 
$\alpha \approx 0.13$ is much smaller than the BCS value 
of $0.5$.  However, for $\lambda=0.6$, i.e.  inside the 
CRO regime, the IE is very large once $\omega_0<0.3t$.  Here the IE coefficient 
is $\alpha \approx 1.41$. 

\begin{figure}[t]
\begin{center}
\includegraphics*[width=3.3in]{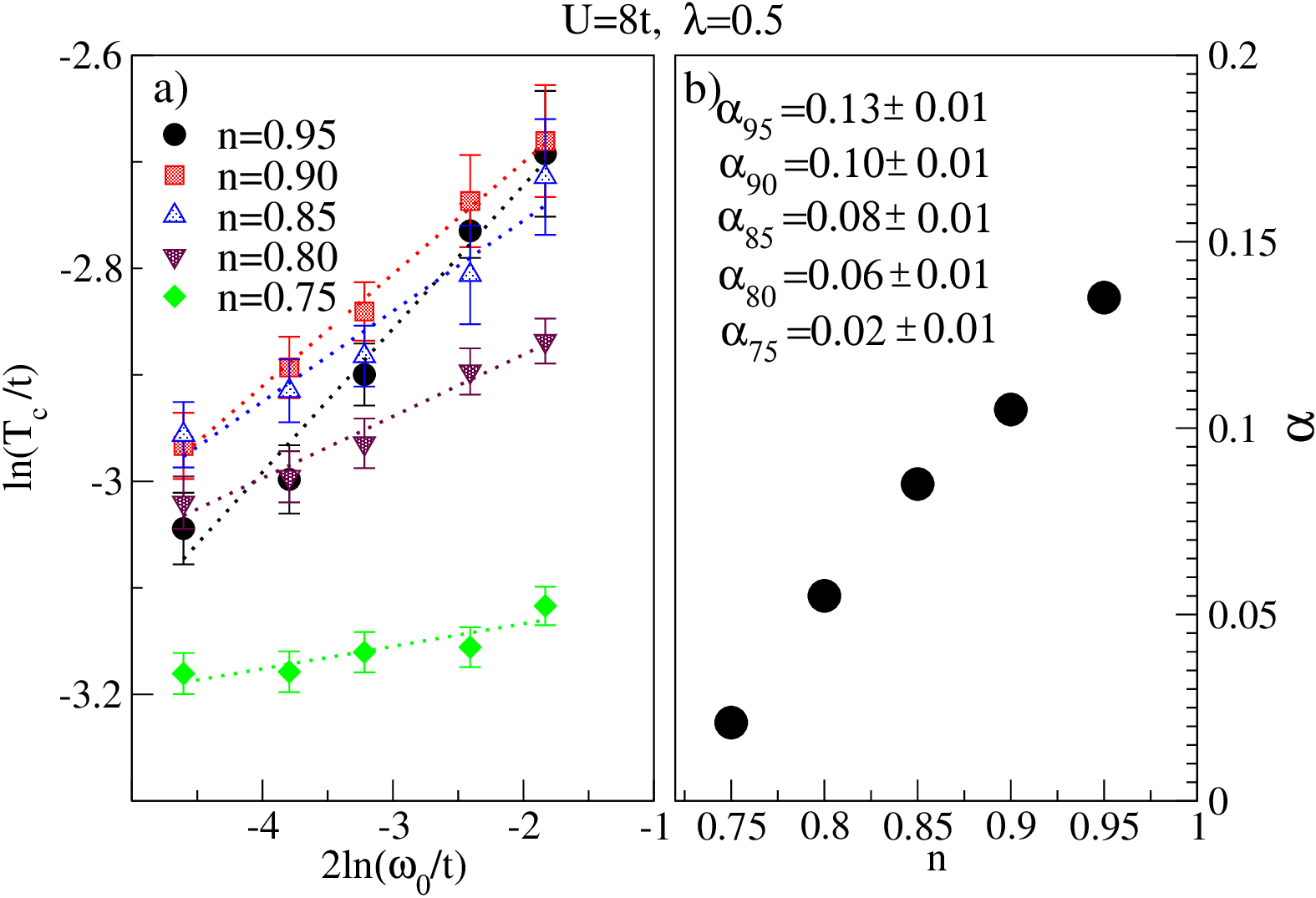}
\caption{(color online) (a) $T_c$ versus  $\omega_0$ at $5\%, 10\%, 15\%$ , $20\%$ 
and $25\%$ doping for  $\lambda=0.5$, on a logarithmic scale. A linear fitting
of $\ln(T_c)$ versus  $2\ln(\omega_0)$ provides an estimation of IE
coefficient $\alpha$ from the slope.  (b) $\alpha$ versus the filling $n$ at $\lambda=0.5$. 
The  error bars are of the order of the symbol size.  $\alpha$  decreases with 
increasing doping.}
\label{fig:ien}
\end{center}
\end{figure}

We find that  the IE is decreasing with increasing   doping.
Fig.~\ref{fig:ien} (a) shows the logarithm of $T_c$ versus the  
logarithm of $\omega_0$ at different dopings 
when  $\lambda=0.5$.  To calculate the IE coefficient we use a linear fitting
of $\ln(T_c)$ versus  $2\,\ln(\omega_0)$ for $\omega_0\le 0.4t$, the slope giving the
estimate of $\alpha$ for this study.  The IE coefficient dependence on doping is shown 
in Fig.~\ref{fig:ien} (b).  The decrease of $\alpha$ with increasing 
doping is in agreement with the notion of significant interplay between 
the AF correlations, EP interaction and superconductivity \cite{macridin_hh}. 
Given stronger AF correlations at smaller doping, the polaronic 
effects become more pronounced and hence the IE is larger. It would be 
interesting to study the doping dependence of the IE for larger values 
of  $\lambda$ (e.g. $\lambda=0.6$).  Unfortunately at large  $\lambda$  
and intermediate doping (i.e. $\approx 15\%$) the charge susceptibility of the system 
becomes extremely large implying strong charge fluctuations that make it
very difficult to stabilize the calculation at low temperature.

\begin{figure}[t]
\begin{center}
\includegraphics*[width=3.3in]{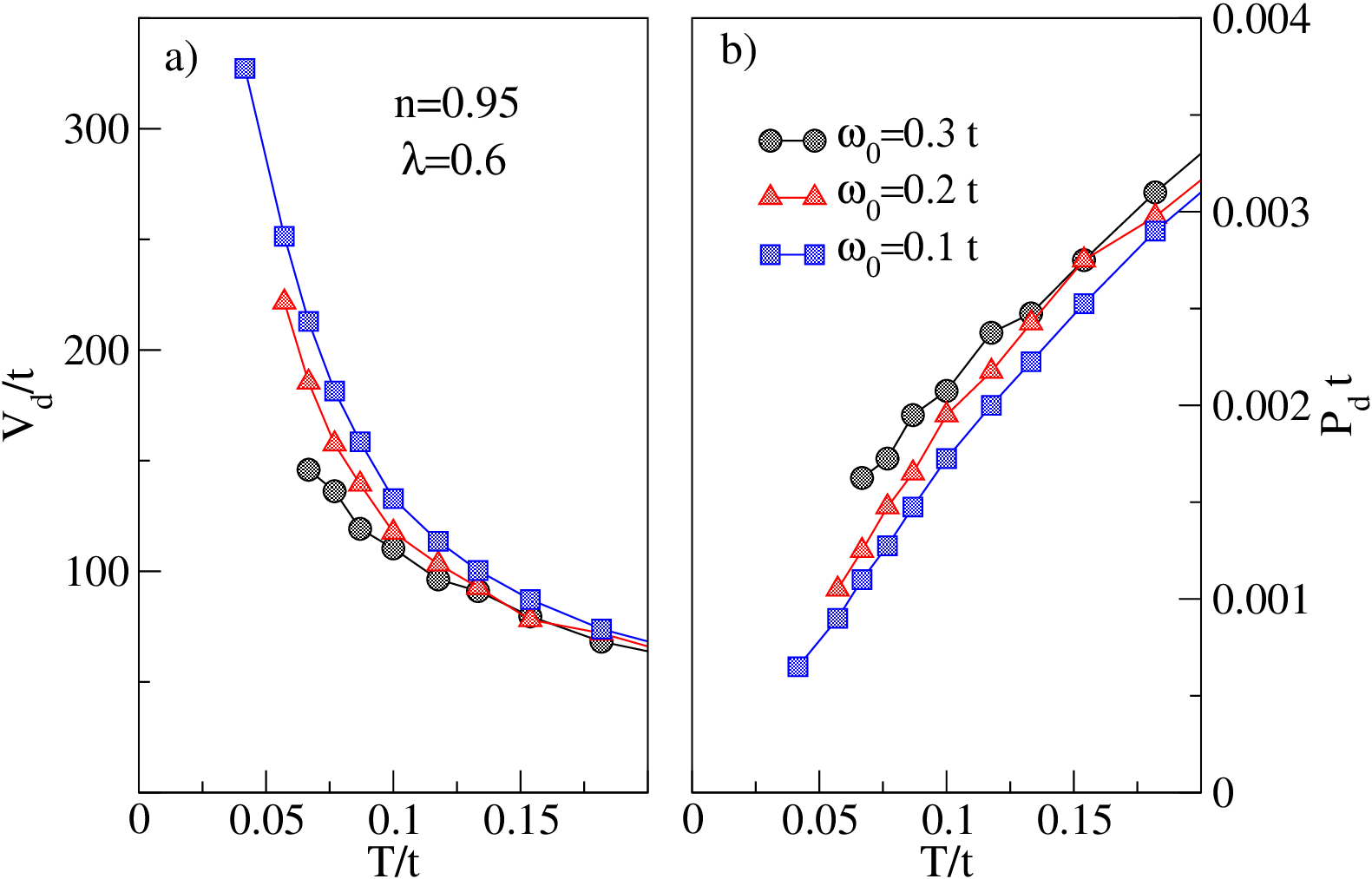}
\caption{(color online) (a) The $d$-wave pairing interaction $V_d$  and 
(b) the $d$-wave projected bubble $P_{d}$
versus temperature for different $\omega_0$, at  $\lambda=0.6$
and $5\%$ doping. While the pairing interaction increases with decreasing $\omega_0$, the decrease of  $P_{d}$
results in an overall suppression of $T_c$ with decreasing $\omega_0$. }
\label{fig:vdpd}
\end{center}
\end{figure}

In regard to superconductivity, the EP coupling gives rise to two competing 
effects \cite{macridin_phonon}, an increase of the effective pairing interaction  
and a renormalization of the single-particle propagator.
The first favors superconductivity while the latter opposes it.  Keeping the
EP coupling fixed and decreasing $\omega_0$ results in a similar 
competition; i.e.,  an increase of the effective pairing interaction 
and a stronger renormalization of the single-particle propagator leading to the 
same overall effect, a reduction of T$_c$.
As discussed in Ref. \cite{maier:pairprb}, the $d$-wave projected 
quantities $V_d$ and $P_d$ provide, respectively, a measure of the effective 
$d$-wave pairing interaction and renormalization of the 
single-particle propagator relevant for $d$-wave superconductivity.
At low temperature as $\omega_0$ decreases $V_d$ increases
while $P_d$ decreases,  as shown in Fig.~\ref{fig:vdpd}.

\begin{figure}[t]
\begin{center}
\includegraphics*[width=3.3in]{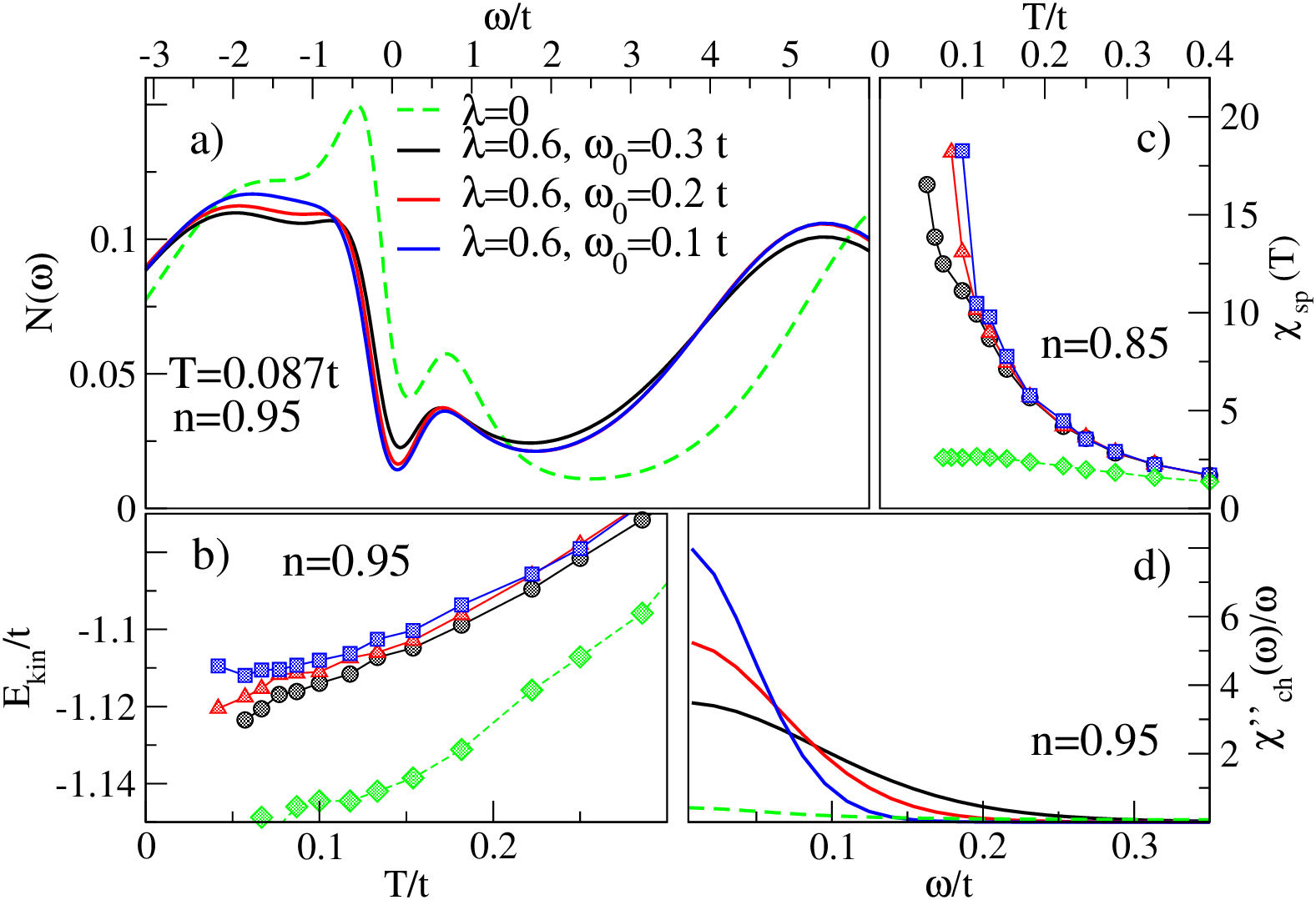}
\caption{(color) The black (red, blue) lines represent $\omega_0=0.3t$ ($\omega_0=0.2t$, $\omega_0=0.1t$) results  for  
$\lambda=0.6$ while the green ones show data for
$\lambda=0$.  (a) DOS $N(\omega)$  and (b) kinetic energy $E_{kin}=\sum_{k,\sigma}\epsilon(k) n_{k,\sigma}$ at  $5\%$ doping. 
The low energy DOS and the kinetic energy gain  are suppressed as $\omega_0$ decreases. 
c) The AF susceptibility $\chi_{sp}$ at  $15\%$ doping.
The AF correlations are enhanced  when $\omega_0$ decreases.
d) The local dynamic charge susceptibility $\chi^{''}_{ch}(r=0,\omega)/\omega$ at  $5\%$ doping.
The width of low energy  peak in the charge excitation spectra is proportional to  $\omega_0$.}
\label{fig:dos}
\end{center}
\end{figure}

In the CRO regime the IE on the superconducting T$_c$ is accompanied by a  
significant IE in the low energy DOS and in the kinetic 
energy of the system. 
For $\lambda=0.6$ when $\omega_0$ decreases, the DOS in an energy 
window of order  $\approx \omega_0$ around the chemical potential 
(chosen zero in Fig.\ref{fig:dos} (a)) is suppressed and the kinetic 
energy at low temperature increases with decreasing $\omega_0$, 
as shown in Fig.~\ref{fig:dos} (b).  Like the projected bubble 
$P_d$ discussed in the previous paragraph,  both the DOS and the kinetic energy 
describe single-particle propreties of the system and are strongly linked
to effects on charge carrier mobility and effective 
mass. Their dependence on $\omega_0$ shows that the polaronic
effects become stronger as the phonon frequency is reduced, 
a fact understood by noting that decreased
$\omega_0$ moves the system closer to the adiabatic regime 
where the CRO region shrinks to $\lambda_c$ and  charge carriers should eventually self-trap 
when $\lambda > \lambda_c$ \cite{adiabatic}.

The magnetic properties are also influenced by the phonon energy.
At small doping and at half filling the magnetic susceptibilities show very small 
$\omega_0$ dependence, on the order of the error bar,  even in the 
CRO regime (not shown). However, the IE on the Ne\'el
temperature $T_N$ can be significant near the doping range where 
$T_N \to 0$.  For example,  as can be seen in Fig.~\ref{fig:dos} (c) at 
$15\%$ doping, the AF susceptibility increases as $\omega_0$ decreases.
The pseudogap temperature $T^*$ measured as the characteristic 
temperature for suppression of low energy spin excitations is virtually 
independent of $\omega_0$. 
It is interesting that when $T^*$ is measured as the temperature  
at which the pseudogap first begins to appear in the DOS, the same 
conclusion can be drawn. Moreover, one can see in Fig.~\ref{fig:dos} (a) 
that the magnitude of the pseudogap measured by the peak-to-peak 
distance increases weakly as $\omega_0$ decreases.

The charge excitation spectrum shows a strong dependence on $\omega_0$.
In the CRO regime the local (i.e. $r=0$) charge excitation spectrum displays a narrow peak 
in agreement with the assumption that the system tends toward formation of a polaronic band 
\cite{macridin_hh}. Moreover, and also in agreement with that assumption,  the width of the 
peak is proportional to  $\omega_0$ as shown in Fig.~\ref{fig:dos} (d).  The 
low-energy weight of the charge spectrum increases as $\omega_0$ decreases and
explains the stronger fluctuations in the electronic density  and the difficulty
in fixing the doping level as $\omega_0$ decreases at low temperatures.

{\em{Discussions.}}
While the HH Hamiltonian shows an IE in many respects similar to the one seen in cuprates,
as we will discuss in the next paragraph,  
a direct comparison with experiments requires  more realistic models which consider
the  symmetry  of the relevant phonon modes.
Since the relevant phonons are the ones associated with oxygen vibrations, the explicit inclusion of 
the oxygen orbitals in the model might be important as well. 

However, we find many similarities between the IE in cuprates and our results in the CRO regime. 
The experiments in cuprates report a  decrease of $\alpha$ with increasing doping \cite{IEcuprates},
a negligible  IE in the pseudogap and AF properties at small doping \cite{ie_pg, ie_af}
and a large IE in the spin glass freezing temperature \cite{zhao_sg} indicative of an 
increase of AF correlations at finite doping with decreasing phonon frequency.
The  large IE seen in the penetration depth
shows strong dependence of the Cooper pairs effective mass on phonon 
frequency \cite{ie_penetration,Khasanov,zhao2}. Although we do not address 
the  superconducting state, this may be understood 
with our findings of strong $\omega_0$ dependence of the charge carriers 
effective mass. The competition between  the enhancement of the 
pairing interaction and the renormalization of the single-particle propagator (discussed in Fig.~\ref{fig:vdpd})
may  be relevant for explaining the negligible  IE in the superconducting T$_c$ accompanied by a  significant IE in the charge carrier effective mass seen in the optimally doped cuprates\cite{zhao2}, 
since the former is a result of the competition  and the latter is a 
consequence of only the  single-particle renormalization.  The experiments find an extremely large IE  around $8\%$ doping reportedly linked with stripe formation \cite{ie_stripe,IEcuprates}.
Stripes cannot be addressed with the small $2 \times 2$ clusters since the the stripe phase order parameter is incommensurate  with the cluster. However, the strong enhancement of charge fluctuations due to EP coupling and the dependence  of charge susceptibility on $\omega_0$ suggest a large IE on stripe formation if larger clusters were considered.

Due to the nature of cluster-mean field approximation which neglects spatial correlations outside the cluster
range, in the underdoped region where the phase fluctuations (PF) are strong, the polaronic effects' influence on the PF is underestimated. This leads to the expectation of an even larger IE at small doping and a stronger doping  dependence of $\alpha$ in  calculations which manage to capture longer range correlations. This can be understood by noting
that the PF coherence temperature is given by the $\frac{n_s}{m^*}$, where $n_s$ is the superconducting
fluid density and  $m^*$ is the effective mass of the pairs \cite{phase_fluc}. Thus 
increased effective mass  due to the decrease in $\omega_0$ amplifies the PF and  reduces $T_c$.


{\em{Conclusions.}}
This letter presents results of calculations of the two-dimensional Hubbard 
model with Holstein phonons using DCA and QMC specifically addressing the IE
in light of recent experiments on high T$_c$ materials, including the cuprates. 
At small  EP coupling there is a negligible IE, but for larger EP values, in the CRO regime,
a large and positive IE effect is found on the superconducting 
T$_c$ that becomes stronger at small doping. A significant IE is seen also in single-particle 
quantities, like the low energy DOS and the kinetic energy, as well as 
the charge excitation spectrum.  This indicates that the IE on T$_c$ is  
directly related with the enhancement of effects such as 
suppression of QP weight and charge carrier mobility
with decreasing $\omega_0$. A larger IE at smaller doping is 
a consequence of an increase in these same effects in the presence of strong 
AF correlations.  A negligible IE is seen in the pseudogap and 
AF properties at small doping whereas the AF 
susceptibility at intermediate doping increases with decreasing phonon 
frequency $\omega_0$.

\acknowledgments  The authors would like to thank Brian Moritz for helpful discussions. This research was 
supported by NSF DMR-0706379 and CMSN DOE DE-FG02-04ER46129.  Supercomputer 
support was provided by Texas Advanced Computing Center.


\begin{thebibliography}{99}
\bibitem{lambdac} $\lambda_c$ and the CRO region are dependent on the phonon mode, e.g. in \cite{macridin_phonon} we find $\lambda_c \approx 0.25$ and a narrower CRO region 
for the buckling mode.
\end{thebibliography}
\end{document}